\documentstyle[aps,psfig,12pt,floats,tighten]{revtex}

\begin{document}
\draft
\title{Self-organized Beating and Swimming of Internally Driven Filaments}
\author{S\'ebastien Camalet, Frank J\"ulicher and Jacques Prost}

\address{Institut Curie, Physicochimie Curie, U.M.R. 168, 26 rue d'Ulm,
75248 Paris Cedex 05, France}

\maketitle

\vskip 0.2in

\begin{abstract}
\hskip 0.15in 
We study a simple two-dimensional model for motion of an elastic
filament subject to internally generated stresses and show that
wave-like propagating shapes which can propel the filament can be
induced by a self-organized mechanism via a dynamic instability. The
resulting patterns of motion do not depend on the microscopic
mechanism of the instability but only of the filament rigidity and
hydrodynamic friction. Our results suggest that simplified systems,
consisting only of molecular motors and filaments could be able to
show beating motion and self-propulsion.
\end{abstract}

\vskip 0.4in

Cilia and flagella are hair-like appendages of many cells which
generate motion and are used for self-propulsion and to stir the
surrounding fluid. They all share the characteristic architecture of
their core structure, the axoneme, a common structural motive that was
developed early in evolution. It is characterized by nine parallel
pairs of microtubules, which are long and rigid protein filaments,
that are arranged in a circular fashion together with a large number
of dynein molecular motors \cite{albe94}. In the presence of a fuel
which is ATP, the dynein motors attached to the microtubules generate
relative forces while acting on neighboring microtubules. The
resulting internal stresses induce the bending and a wave-like motion
of the axoneme.

These biological systems are complex, they consist of a large number
of different components and various patterns of motion have been
observed.  Attempts to model their behavior are either based on the
assumption that some unknown control system generates
oscillatory motor activity \cite{control} or that a self-organized
mechanism is at work \cite{brok75,lind95}.  Generically, the
latter involves a dynamical instability.
Theoretical studies of simple models for collective action of
molecular motors have demonstrated the possibility of such
instabilities \cite{brok75,juli95,juli97,hang98}.
 Several examples of oscillatory motion of biological
many-motor systems are known. Recently, it was suggested that
spontaneous oscillations observed in muscles could be a property of
the motor-filament system alone \cite{juli97,fuji98}. This idea is
supported by the fact that the oscillations continue to exist after
all regulatory systems are removed \cite{fuji98} but also by the
observation that an in vitro motor-filament system shows the signature
of a dynamic transition
\cite{rive98}. Furthermore, the observations that flagellar dyneins
are able to generate oscillatory motion on microtubules
\cite{shin98}, and that isolated and de-membranated flagella in
solution containing ATP 
above a threshold concentration swim with a
simple wave-like motion
\cite{gibb75}
support the idea that basic types of flagellar beating could result
from a dynamic instability.

\begin{figure}
\centerline{\psfig{figure=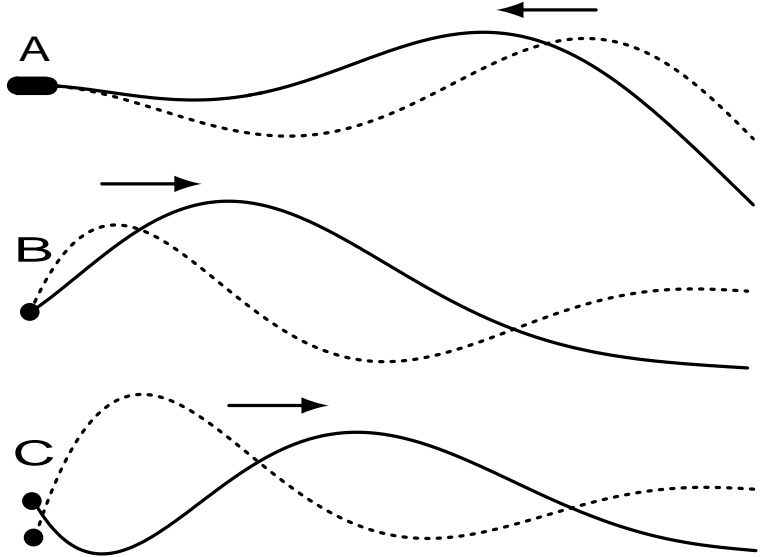,width=10.0cm}}
\bigskip
\caption{Snapshots of wave-like patterns generated by a motor-induced
Hopf-bifurcation calculated for different boundary conditions (solid
lines): (A) Clamped head, position and slope are fixed. (B) Fixed
head, position is fixed only.  (C) Free head subject to a viscous
load. The broken lines represent earlier configurations.  The arrows
indicate the direction of wave propagation.}
\label{f:snapshots}
\end{figure}

In this article, we introduce a simple two-dimensional model which
reveals many physical aspects of the motion of an elastic filament driven
by internal forces, that should be relevant for flagellar
beating.  Our approach is inspired by studies of
semiflexible filaments subject to external forces
\cite{ever95,gold95,bour95,seki95,wigg98}, 
however, in our case all motion is induced by {\em internal stresses}.
Our model consists of two incompressible but elastic filaments of
length $L$ arranged at constant distance $a\ll L$ and rigidly attached
only at one end which we call the head.  A large number of molecular
motors and passive elements holding the filament pair together are
assumed to generate a coarse-grained force per unit length $f$ which
is an internal stress acting in opposite directions on the two
filaments and induces the bending of the filament pair. The dynamic
equations of this model define patterns of beating motion resulting
from the internal forces which are assumed to oscillate. More
interestingly, we show that characteristic wave-like patterns which
propagate along the filament are generated most naturally by a dynamic
instability of the motor-filament system, see Fig. \ref{f:snapshots} for
examples. As we show below, the qualitative shapes of these patterns
do not depend on the microscopic mechanism of force generation but
only on the elastic properties of the filaments and on hydrodynamic
friction. We demonstrate that these patterns lead to self-propulsion
of the system and calculate the velocity of motion.

In order to define our model and to derive the dynamic equations, we
start from the enthalpy functional
\begin{equation}
G=\int_0^{L}  \left[ \frac{\kappa}{2}(C(s))^2 + f(s) \Delta(s) + 
\Lambda(s)(\partial_s \vec r)^2 \right ] ds
\label{eq:Fu}
\end{equation} 
where $\vec r(s)$ is a parameterization of the shape of the filament
pair by the arclength $s$, $\kappa$ is the bending rigidity, and
$C=\vec n \cdot \partial_s^2\vec r$ is the local curvature with the
filament normal $\vec n$.  The internal force
density $f$ couples to the relative local displacement of the filament
pair $\Delta(s)= a \int_0^{s} C(s') ds'$ \cite{ever95}.  In order to impose the
constraint of local incompressibility $(\partial_s \vec r)^2=1$, we
have introduced the Lagrange multiplier $\Lambda(s)$. The equation of
motion can be written as
\begin{equation}
\partial_t \vec r = -\left(\frac{1}{\xi_{\perp}} \vec n \vec n +
 \frac{1}{\xi_{\parallel}} \vec t\, \vec t\,\right) 
{\frac{\delta G}{\delta \vec r}}
\label{eq:teu}
\end{equation} 
where $\vec t\vec t$ and $\vec n\vec n$ are projectors on the filament
tangent and normal, and we assume local anisotropic friction with
tangent and normal coefficients $\xi_{\parallel}$ and
$\xi_{\perp}$, respectively. The Lagrange multiplier is determined by
the condition $(\partial_s\vec r)\partial_t\partial_s \vec r=0$
\cite{gold95}.

In order to keep the description simple, we consider small
deformations of a filament parallel to the $x$-axis, $\vec
r(s)=(s+u(s),h(s))$, which we describe by an expansion in the
transverse and longitudinal displacements $h$ and $u$. To quadratic
order in $\partial_x h(x)$ we can write
\begin{equation}
G\simeq \int_0^L\left[\frac{\kappa}{2} (\partial_x^2 h)^2 
+ a f(x)(\partial_x h(x) - \partial_x h(0))\right ]dx
\label{eq:F} \quad ,
\end{equation}
where we use the $x$-coordinate as parameter.  We first discuss
transverse motion which for small deformations is independent of
longitudinal forces \cite{fnlam} and satisfies the equation
$\xi_{\perp} \partial_t h = -
\kappa \partial_x^4 h + a \partial_x f$ together with two
boundary conditions at the head with $x=0$ and two conditions at the
tail for $x=L$.  We assume a free tail which implies
$\partial_x^2 h(L) = 0$ and $\kappa
\partial_x^3 h (L) = a f(L)$. At the head, we distinguish three
different cases as shown in Fig. \ref{f:snapshots}: (A) clamped head with
$h(0)=0$ and $\partial_x h(0)=0$; (B) fixed head with
$h(0)=0$ and $\kappa
\partial_x^2 h (0) =- a \int_0^L f(x) dx$; and (C) a viscous load at $x=0$ with
friction coefficient $\eta$ for which the condition on $h(0)$ in (B)
is replaced by $\eta \partial_t h(0) =a f(0) -\kappa \partial_x^3 h
(0)$.

\begin{figure}
\centerline{\psfig{figure=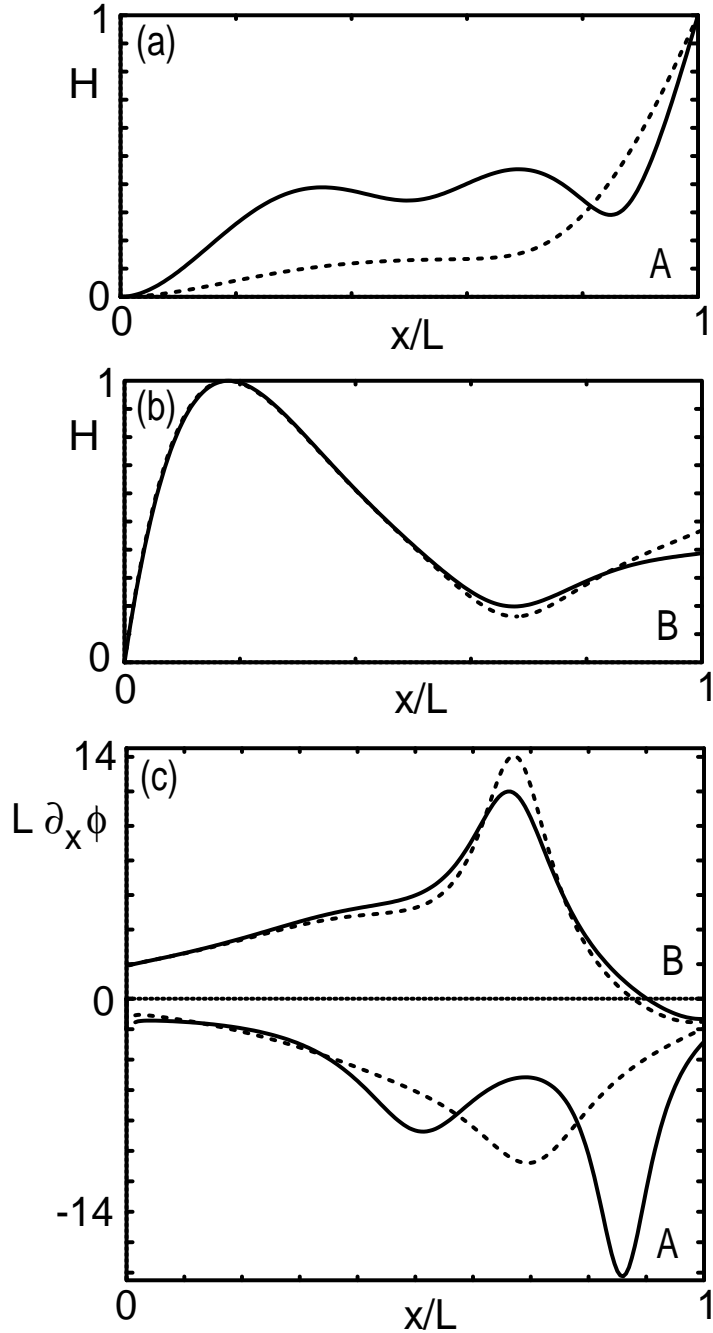,width=10.0cm}}
\bigskip
\caption{(a) Amplitude $H(x)$ (in arbitrary units) of the wave-like 
motion characterized by Eq. (\protect\ref{eq:wave}) as a function of
the position $x$ along the filament axis for boundary conditions (A)
as defined in Fig. \ref{f:snapshots} and $\xi_{\perp} \omega L^4/\kappa =
2500$. (b) same plot for boundary conditions (B).  (c) Gradient
$\partial_x \phi$ of the phase along the filament axis for the same
systems.  The solid lines correspond to motion induced by a
Hopf-bifurcation for the smallest response coefficient $\chi_1$, the
broken lines to motion induced by a homogeneous internal force and
$\chi=0$.}
\label{f:waves}
\end{figure}

We demonstrate the basic properties of this model, by first assuming
that an oscillating force density with constant amplitude is generated
by some unspecified mechanism: $f_m(t)={\rm Re}(\tilde f_0 e^{i\omega
t})$.  The total force density $f$ acting on the filament pair is the
sum of the force $f_m$, internal dissipative forces and in general the
forces of elastic elements which locally connect the filaments.
Introducing the complex Fourier amplitude $\tilde h$, where
$h(x,t)={\rm Re} (\tilde h(x) e^{i\omega t})$, we can express the total
force density as $\tilde f = \chi \tilde v + \tilde f_0$, where $\tilde
v\simeq i\omega a (\partial_x \tilde h(x) - \partial_x \tilde h(0))$ is
the complex amplitude of the local sliding velocity
$v=\partial_t\Delta$. The coefficient $\chi=(\lambda+K/i \omega)$
describes a viscoelastic response of the material between the
filaments with dissipation coefficient $\lambda$ and elastic
modulus $K$. The oscillating state is characterized by
\begin{equation}
\kappa  \partial_x^4 \tilde h - a^2 i \omega \chi \partial_x^2 
\tilde h + \xi_{\perp} i \omega \tilde h = 0
\label{eq:ins} \quad .
\end{equation}
The homogeneous active force $\tilde f_0$ only enters via boundary
conditions. Eq. (\ref{eq:ins}) and boundary conditions represent an
inhomogeneous linear system which is solved by $\tilde h=A e^{k x/L}$
leading to four complex values of $k$. The corresponding coefficients
$A$ are adjusted to satisfy the boundary conditions which leads to a
solution with an amplitude proportional to the internal force $\tilde
h(x) \sim \tilde f_0$. We can distinguish two different regimes: 
(i) hydrodynamic
friction dominates $|\chi|^2 \ll \kappa \xi_{\perp} /\omega a^4$;
(ii) internal viscoelasticity dominates
$|\chi|^2 \gg \kappa \xi_{\perp} /\omega a^4$. We can neglect in 
Eq. (\ref{eq:ins}), $\chi$ in case (i) and $\xi_\perp$ in case (ii).
Fig. \ref{f:waves} shows examples of the amplitude $H$ and the
gradient of the phase $\phi$ of $\tilde h(x)=H(x)e^{-i\phi(x)}$ for
$\chi=0$ and different boundary conditions as dashed lines. The
corresponding time dependent solutions
\begin{equation}
h(x,t)=H(x)\cos(\omega t -\phi(x)) \label{eq:wave}
\end{equation} 
are propagating wave-like shapes qualitatively similar to those shown
in Fig. \ref{f:snapshots}. The sign of the local propagation velocity
$v_p=\omega/\partial_x\phi$ of the phase allows us to determine
the direction of apparent wave propagation.

We have thus developed the framework to calculate and analyze
wave-propagating solutions of our model and can now study motion
generated by the properties of the motor-filament system via a Hopf
bifurcation. We assume that the material between the two filaments
which contains both molecular motors and passive elements has
properties which can be characterized on a coarse-grained level by a
nonlinear history-dependent response function. We will study the
instability of a non-moving solution $h(x)=0$ towards wave-like
patterns.  For this case it is sufficient to consider only small
amplitudes, $|\partial_x h|\ll 1$ as described above.  Furthermore, in
this regime the local sliding velocity $v$ is small and we can ignore
nonlinearities in $v$ and restrict ourselves to the frequency
dependent linear response $\tilde f = \chi \tilde v$.  Here, we have
set the artificially introduced force $\tilde f_0=0$ and characterize
both passive and active material properties by the complex response
function $\chi(\omega,\Omega)$ which can e.g. be calculated explicitly
for a simple model
\cite{juli97} or measured experimentally \cite{mach59}.  The out-of-equilibrium nature of the
system is characterized by the parameter $\Omega$ which can for
example be identified with the ATP concentration.  Note, that for an
active system $\chi$ can have unusual behaviors which formally
correspond to a negative friction (${\rm Re}(\chi)<0$) or a negative 
elastic response (${\rm Im}(\chi)>0$).

In the case $\tilde f_0=0$, Eq. (\ref{eq:ins}) and boundary conditions
becomes a homogeneous linear system which always has the solution
$\tilde h(x)=0$ and which can now be reinterpreted as an eigenvalue
problem for $\chi$.  Spontaneous motion corresponds to nontrivial
solutions to this problem. A discrete set of such solutions $\tilde
h_i$ exists, each $\tilde h_i$ corresponds to a complex eigenvalue
$\chi=\bar\chi_i(\omega)$, $i=1,2\dots \infty$ which we order
according to $|\bar\chi_i(\omega)|\leq|\bar\chi_{i+1}(\omega)|$.

Consider now a system initially at equilibrium with $\Omega=0$.  If $\Omega$
is increased, an instability occurs as soon as a critical value
$\Omega_c$ is reached for which
$\chi(\omega_c,\Omega_c)=\bar\chi_i(\omega_c)$ for a frequency
$\omega_c$. In the vicinity of this point, the system develops for
$\Omega>\Omega_c$ motion with this frequency and a shape characterized
by the nontrivial solution $\tilde h_i(x)$. This scenario applies to a
supercritical bifurcation. Nonlinear terms of the response function
and nonlinear corrections to the simple Monge representation can
become important for larger $\Omega$, or they could change the
nature of the bifurcation to subcritical.
 Typically, the instability occurs for the
smallest value $\chi=\bar\chi_1(\omega)$ since larger $|\chi|$ require
larger values of $\Omega$ which correspond to more system
activity. Note, that the resulting pattern of motion is independent of
the microscopic mechanism which leads to the instability. It is
sufficient that the active material is capable to generate the
response $\chi=\bar\chi_i$ \cite{fninst}.

Fig. \ref{f:waves} displays examples of the amplitude and the
gradient of the phase of $\tilde h_1(x)$ for boundary conditions (A)
and (B), snapshots of the corresponding motion are shown in
Fig. \ref{f:snapshots}. The boundary conditions play an essential role in
selecting different types of motion. Observing the sign of
$\partial_x\phi$ which determines the direction of the phase velocity
we find that for clamped head (A) the wave propagates from the tail
towards the head while for case (B) it propagates in the opposite
direction.  The amplitude $H(x)$ also differs significantly between
cases (A) and (B), see Fig. \ref{f:waves}. The case (C) of a free
head with viscous load $\eta$ is similar to case (B) and therefore not
shown in Fig. \ref{f:waves}. For this example, the qualitative
properties of motion induced by the dynamic instability are the same
as those of the system driven by a homogeneous force $\tilde f_0$, see
Fig. \ref{f:waves}.  In fact, for the parameters chosen,
$|\chi_1|^2 \ll\kappa \xi_{\perp} / (\omega a^4)$ and the corresponding
solution is not far from the solution for $\chi=0$.  The case of
homogeneous force $\tilde f_0$ is simple and allows us to explain the
effect of boundary conditions. A homogeneous internal force $\tilde
f_0$ can be rewritten as boundary terms in the expression of the
energy: $G\simeq af_0h(L)-af_0h(0)-aLf_0\partial_x h(0)+\int_0^L
dx\,(\partial_x^2 h)^2
\kappa/2$.  Its action is equivalent to two opposite transverse forces
$af_0$ acting at both ends together with a torque $aLf_0$ applied at
the head. In the case of a clamped head this apparent force and torque
are suppressed and the system is driven by a virtual force at the
tail, propagating the wave towards the head \cite{wigg98}. If the head
is not clamped, the virtual oscillating torque at the head can
propagate a wave in the opposite direction.

Can these beating patterns propel the filament?  Time-reversal
symmetry has to be broken, $h(x,-t) \ne h(x,t)$, for propulsion
to be possible \cite{purc77}. According to Eq. (\ref{eq:wave}), this
requirement is fulfilled since $\partial_x
\phi\neq 0$. Because of the symmetry $h(x,t)=-h(x,t+\pi/\omega)$,
there can be no net motion in the transverse direction.
 In order to estimate longitudinal motion, we have to 
study the displacement $u(x)$. To
second order in $\partial_x h$, we can write
\begin{equation} 
 u(x) \simeq u(0) -\frac{1}{2} \int_0^x (\partial_{x} h)^2 dx' \quad ,
\label{eq:long}
\end{equation}   
indicating that the dynamics of $u(x)$ is governed by the motion
$h(x,t)$. Note that $u(x)-u(0)$ is small, but the filament
displacement $u(0)$ can become large. 
The longitudinal component $f_l$
of the hydrodynamic force density $ -(\xi_{\perp} \vec n \vec n +
\xi_{\parallel} \vec t\, \vec t\,)\cdot\partial_t \vec r$ acting locally
on the filament is given by $f_l\simeq(\xi_{\perp} -\xi_{\parallel})\partial_x h 
\partial_t h-\xi_{\parallel} \partial_t u(x,t)$ 
in our approximation. The velocity of motion $V$ is the time-average
of $\partial_t u(0)$ and follows from the condition that the total
longitudinal force vanishes. If an isotropic 
viscous load is attached to the head, this condition is 
$\int_0^L f_l dx+\eta \partial_t u(0)=0$ and we find 
$V=V_0/(1+\eta/\xi_{\parallel}L)$ where
\begin{equation}
V_0 = - \left(\frac{\xi_{\perp}}{\xi_{\parallel}}-1\right) \frac{\omega}{2L}
 \int_0^L H(x)^2 \partial_x \phi\; dx
\label{eq:v0}
\end{equation} 
is the no-load velocity. If the head is not permitted to move, the
filament generates a force $F=V_0\xi_\parallel L$ at the head. 
Note, that for isotropic friction
both $V$ and $F$ vanish. For a semiflexible rod-like filament
$\xi_\perp/\xi_\parallel\simeq 2$
\cite{doi86} and the direction of motion is opposite to the direction
of phase propagation.

The parameters chosen in Figs. \ref{f:snapshots} and \ref{f:waves}
correspond e.g. to $L\simeq 40\mu$m, $a\simeq 20$nm,
$\xi_{\perp}\simeq 2\cdot 10^{-3} $Ns/m$^2$, $\kappa\simeq 4\cdot
10^{-22}$Nm$^2$, which is the elastic modulus of about $20$
Microtubules \cite{gitt93}, and a frequency $\omega/(2 \pi)\simeq
30\,$s$^{-1}$.  For this choice, we find a critical value
$\chi_1\simeq (-10+20i)$Ns/m$^2$.  We choose an amplitude of $\tilde
h_1$ with maximal value $H/L\simeq 0.1$. In this case, the maximal
local sliding velocity is $v\simeq 8 \mu$m/s. In axoneme, dynein motors are
spaced every $24$nm along the microtubules. Assuming that only one
microtubule pair is active, we estimate that for this choice a force
per motor of $4$pN corresponds to the critical value $f\simeq
|\chi_1|v$. This is a typical force created by molecular motors.
Larger forces could be necessary to generate beating with larger
amplitudes. Our result suggests that in this case several microtubule
 pairs could be active at the same time thus allowing for
smaller forces per motor.  Using the motion $\tilde h_1$ obtained for
boundary conditions (C) with a viscous load $\eta=
5\cdot10^{-8}Ns/m$, we find a spontaneous velocity of lateral motion
$V\simeq 40\mu$m/s, which is significantly larger than local sliding
velocities and not far from experimentally observed values.

We have demonstrated that a pair of elastic filaments held together by
an active, force-generating material, can induce wave-like patterns by
a dynamic instability of the system.  This study is motivated by
biological flagella such as those of sperms which use such motion for
self-propulsion.  Our model suggests that the boundary conditions
imposed at the ends select the type of beating pattern. This
could be tested by micro-manipulation experiments which apply external
forces and torques at the ends of beating flagella.  We have
restricted our study to a two-dimensional system, small deformations
and the linear regime of the instability.  Nonlinearities could play an
important role in the regime of rapid self-propulsion with large
amplitudes.  Furthermore, the possibility of torsional motion in
three-dimensional systems could allow for new types of behavior.

The observation that wave-like motion can be generated in a
self-organized way raises the idea that sophisticated control
mechanisms may have evolved after the development of the basic
axonemal structure in order to fine-tune the system and to create more
complex types of motion. This concept suggests that artificially
constructed systems consisting only of motors and filaments could
already undergo beating motion and self-propulsion.

We thank R. Everaers for stimulating collaborations, M. Bornens and
H. Delacroix for introducing us to cilia and flagella, A. Parmeggiani
and C. Wiggins for useful discussions and A. Maggs for a critical
reading of the manuscript.


\begin{references}
\bibitem{fninst} The model discussed in Ref.
\cite{juli97} can induce this instability.
\end{references}
\end{document}